\documentclass[prb,preprint,preprintnumbers,amsmath,amssymb,altaffilletter]{revtex4}
\usepackage{graphicx,epstopdf,subfigure,natbib,amsmath,amssymb}
\begin{document}
\title{Landau thermodynamic potential for BaTiO$_3$}
\author{Y. L. Wang,\footnote{$^)$ Email: yongli.wang@epfl.ch}$^)$ A. K. Tagantsev, D. Damjanovic, and N. Setter}

\affiliation{Ceramics Laboratory, EPFL-Swiss Federal Institute of Technology,
Lausanne 1015, Switzerland}
\author{V. K. Yarmarkin}
\affiliation{A.F. Ioffe Physical-Technical Institute, St.Petersburg,194021 Russia}
\author{A. I. Sokolov}
\affiliation{Department of Quantum Electronics,
Saint Petersburg Electrotechnical University,
Professor Popov Str. 5,
Saint Petersburg, 197376,
Russia}
\author{I. A. Lukyanchuk}
\affiliation{Laboratoire de Physique de la Mati\`ere Condens\'ee,
Universit\'e de Picardie Jules Verne
33 rue Saint-Leu
80039 Amiens Cedex, France}

\date{\today}

\begin{abstract}
In this paper the description of the dielectric and ferroelectric properties of BaTiO$_3$ single crystals using Landau phenomenological thermodynamic potential was addressed. Our results suggest that when using the sixth-power free energy expansion of the thermodynamic potential, remarkably different values of the fourth-power coefficient, $\beta$ (the coefficient of $P^4_i$ terms), are required to adequately reproduce the nonlinear dielectric behavior of the paraelectric phase and the electric field induced ferroelectric phase ($E_b\|$[001]), respectively. In contrast, the eighth-power expansion with a common set of coefficients enables a good description for both phases at the same time. These features, together with the data available in literature, strongly attest to the necessity of the eighth-power terms in Landau thermodynamic potential of BaTiO$_3$. In addition, the fourth-power coefficients, $\beta$ and $\xi$ (the coefficient of $P^2_iP^2_j$ terms), were evaluated from the nonlinear dielectric responses along [001], [011], and [111] orientations in the paraelectric phase. Appreciable temperature dependence was evidenced for both coefficients above $T_C$. Further analysis on the linear dielectric response of the single domain crystal in the tetragonal phase demonstrated that temperature dependent anharmonic coefficients are also necessary for an adequate description of the dielectric behavior in the ferroelectric phase below $T_C$. As a consequence, an eighth-power thermodynamic potential, with some of the anharmonic coefficients being temperature dependent, was proposed and compared with the existing potentials. In general the potential proposed in this work exhibits a higher quality in reproducing the dielectric and ferroelectric properties of this prototypic ferroelectric substance.
\end{abstract}

\maketitle

\newpage
\section{INTRODUCTION}

Structural phase transformation phenomena are an important issue in solid state physics.
Both phenomenological and microscopical approaches have been playing comparable roles in understanding the phase transition nature of the magnetics, superconductors, and ferroelectrics.
Particularly for ferroelectrics, considerable attention is paid to the phenomenological Landau-Devonshire theory in parallel with advancing techniques of first-principle calculations.
In the phenomenology of ferroelectrics, the issue recently attracting attention was the role of high-order anharmonic polarization terms in the Landau free-energy expansion. \cite{Vanderbilt2001,Gufan2002}
In these works, the terms containing eighth (or higher) power of polarization, traditionally neglected in the phenomenological framework, was demonstrated to be vitally important for the description of the phase diagrams of perovskite ferroelectrics (e.g. for the recently discovered monoclinic phase of some complex perovskites in the vicinity of the morphotropic phase boundary).\cite{Noheda1999}
The physics behind this effect is the need of high-order terms for adequate description of the symmetry of the problem.
Another  demarche in the field has been recently undertaken by Li \textit{et al.}, who have revisited the phenomenological theory of classical ferroelectric BaTiO$_3$.\cite{Li2005}
These authors suggested incorporation of the eighth power terms in the Landau expansion of BaTiO$_3$ in order to eliminate the temperature dependence of the anharmonic coefficients and to adapt the description to the situation of large compressive strains.
In contrast to Refs. \onlinecite{Vanderbilt2001, Gufan2002}, here no symmetry arguments have been involved.
The message by Li \textit{et al.} was that the Landau expansion  containing eighth power terms (with temperature independent anharmonic coefficients) is as efficient in the description of many properties of BaTiO$_3$  as the traditional sixth order expansion with the temperature dependent anharmonic terms.\cite{Bell1984, Bell2001}
This conjecture is of the interest since the strong temperature dependence of the anharmonic coefficients in the BaTiO$_3$ Landau expansion is in conflict with the displacive nature of ferroelectricity in the material, at least under the common assumption that the critical fluctuations are weak in this system.\cite{Vaks1970, Sokolov2002}
Interestingly, in their analysis of the Landau potential using first-principles calculations, Iniguez \textit{et al.} obtained that the sixth order expansion accounts for the main features of BaTiO$_3$ phase diagram, but that all coefficients in the expansion have nontrivial temperature dependences.\cite{Iniguez2001}

All in all, these results pose two questions. First, are higher than sixth power terms in the BaTiO$_3$ Landau expansion really needed for the correct physical description and are not just a matter of convenience? Second, does the incorporation of such terms enable elimination of the strong temperature dependence of the anharmonic coefficients in the BaTiO$_3$ Landau expansion? Addressing these questions, the present paper demonstrates by analyzing the dielectric response that, both the eighth-power terms and the temperature dependence of the anharmonic coefficients (at least of the quartic coefficients) are essential for the adequate description of the thermodynamic behavior of BaTiO$_3$, especially on crossing the cubic-tetragonal phase transition. Including these necessities, an improved Landau phenomenological potential is proposed and compared with the existing models. Preliminary results on this subject have been published elsewhere.\cite{Wang2006}

\section{IMPORTANCE OF THE EIGHTH-POWER TERMS}

 \subsection{The effect of the eighth-power terms on dielectric nonlinearity}
Besides the phase stability, the nonlinear dielectric behavior is another characteristic that directly evidences the contributions of the anharmonic terms in Landau thermodynamic potential.
To demonstrate the crucial role of the eighth-power term in the Landau expansion we firstly address a simple case, namely the dielectric nonlinearity of BaTiO$_3$ with respect to the electric field $E$ applied along the $[001]$ axis, in the vicinity of the cubic/tetragonal phase transition temperature, $T_C$.
In this case only the $[001]$ component of the polarization, $P=P_3$, is involved, so that the Landau expansion of the Gibbs potential containing the eighth-power term reads
\begin{equation}
\label{Gibbs8}
\triangle G=    \frac{1}{2} \alpha P^{2}+
            \frac{1}{4} \beta P^{4}+
            \frac{1}{6} \gamma P^{6}+
            \frac{1}{8} \delta P^{8}
            -EP
\end{equation}
Accordingly we have for the equation of state and $c$-axis permittivity $\epsilon_c$:
\begin{equation}
\label{E8}
E=           \alpha P+
             \beta P^{3}+
             \gamma P^{5}+
             \delta P^{7}
\end{equation}
\begin{equation}
\label{lambda8}
\epsilon_c^{-1}=
             \alpha +
            3 \beta P^{2}+
            5 \gamma P^{4}+
            7 \delta P^{6}
\end{equation}
These equations provide a description of the non-linear dielectric response of BaTiO$_3$ in terms of the eighth-power expansion.
One can compare this description with that in terms of the sixth-power expansion
\begin{equation}
\label{E6}
E=           \alpha P+
             \beta^{\prime} P^{3}+
             \gamma^{\prime} P^{5}
\end{equation}
\begin{equation}
\label{lambda6}
\epsilon_c^{-1}=
             \alpha +
            3 \beta^{\prime} P^{2}+
            5 \gamma^{\prime} P^{4}
\end{equation}
which have been traditionally used in the field.
These descriptions are clearly different.
Formally, as one can readily check, the situation described by the expansion Eq. \eqref{E8} and \eqref{lambda8} with the polarization-independent coefficients $\beta, \gamma$, and $\delta$ corresponds to  polarization-dependent coefficients $\beta^{\prime}$ and $\gamma^{\prime}$ in Eq. \eqref{E6} and \eqref{lambda6}:
\begin{equation}
\label{beta'}
\beta^{\prime} =\beta-\delta P^{4}, \gamma^{\prime} =\gamma+2\delta P^{2}
\end{equation}
In general, the consideration of the polarization-dependent anharmonic coefficients in the Landau expansion has no physical sense.
However, if a not too wide interval of polarization variation is experimentally addressed with a finite measurement accuracy, coefficients $\beta^{\prime}$ and $\gamma^{\prime}$ defined by Eq. \eqref{beta'} can be considered as polarization-independent when fitting the dielectric non-linearity.
At the same time, serious problems may arise in situations when a large variation of polarization is involved, e.g. the ``jump'' of the polarization on crossing the first order phase transition.
To verify this point we consider the case just above $T_C$, where the ferroelectric phase can be induced by an electrical field.
As is schematized by the $E\geqslant0$ segment of the ``double hysteresis loop'' in Fig. \ref{double hysteresis loop}, the polarization ``jumps'' at the critical field $E_C$, whereas below or above this value the polarization varies slowly with bias field.
At electric fields corresponding to the paraelectric phase, the polarization varies from zero to the critical value $P_C$. Consequently the contributions of the polarization dependent terms can be estimated from Eq. \eqref{beta'} to be not more than $\delta P_C^4$ and $2 \delta P_C^2$ to $\beta^{\prime}$ and $\gamma^{\prime}$, respectively.
In contrast, at electric fields corresponding to the ferroelectric phase, the polarization ranges in a much higher interval of not less than $P_0$, the minimum value of polarization in ferroelectric phase. Accordingly the polarization dependent correction terms in Eq. \eqref{beta'} will be expected to be not less than $\delta P_0^4$ and $2 \delta P_0^2$, respectively.
In Table \ref{tab1}, we list the numerical values of the polarization dependent terms in both phases and the typical values of the corresponding coefficients. The value of $\delta$ follows Li \textit{et al.}'s suggestion.\cite{Li2005}
Apparently the values of the polarization dependent terms strongly depend on the interval over which the polarization varies.
In ferroelectric phase, the values of the ``correction'' terms are comparable to the typical values of the corresponding coefficients, whereas they are negligible in paraelectric phase.

Thus from the above analysis, we conclude that if the coefficients of the eighth-power terms have a value of the order of that suggested by Li \textit{et al.}, the anharmonic polarization coefficients $\beta^{\prime}$ and $\gamma^{\prime}$ of BaTiO$_3$ estimated in terms of the sixth-power expansion should exhibit appreciable jumps on crossing the field induced phase transition.
Such ``phase sensitivity'' of the nonlinear coefficients of the sixth-power expansion enables the detection of the eighth-power terms, suggesting a simple and direct way to assess the role of the eighth-power terms.

Keeping this in mind, we studied the nonlinear dielectric behavior of BaTiO$_3$ single crystals in paraelectric phase and in electric field induced ferroelectric phase ($E_b//[001]$) at a temperature a few degrees higher than $T_C$.
The experimental details can be found in our previous work.\cite{Wang2006}
Figure \ref{fig2} shows the experimental data measured at 135$^\circ C$ (for this crystal, the phase transition temperature $T_C=130^\circ C$). Little hysteresis effect related to the relaxation phenomena is found within either phase, indicating the high insulating quality of the crystal and the space charge free state of the crystal-electrode interface.\cite{Triebwasser1960}
The sixth-order expansions, Eqs. \eqref{E6} and \eqref{lambda6}, are used to fit the curves with the common value of $\alpha$, and remarkably different $\beta$ are obtained for either phase, as listed in Table \ref{tab2}.
As for $\gamma$, we cannot determine it exclusively from the nonlinear dielectric behavior of the paraelectric phase. The position of $E_C$, where the ferroelectric phase is induced by the bias field, is used to obtain an estimate, which differs considerably from that derived in the ferroelectric phase.
The ferroelectric-phase coefficients have a serious problem in describing the dielectric properties of the paraelectric phase, and vice versa.
As illustrated in the inset of Fig. \ref{fig2}, the eighth order expansion with a common set of coefficients enables a good fit in both phases at the same time.
As-estimated values of $\beta$, $\gamma$, and $\delta$, agree well with the set suggested by Li \textit{et al.}\cite{Li2005}

Furthermore this approach is extended to a wider temperature interval around $T_C$. As shown in Fig. \ref{fig3}, the quartic coefficient determined with the sixth-power expansion exhibits strong phase dependence, characterized by a ``jump'' on crossing the phase transition no matter whether this transition is induced by electric field or temperature.
In contrast the values determined with the eighth-power expansion are phase independent as Landau theory coefficients must behave.
The ``phase sensitivity'' of $\beta$ determined by the sixth-power expansion, and its elimination by the incorporation of the eighth-power term, strongly attest to the essentiality of the eighth-power terms in BaTiO$_3$ Landau expansion.

\subsection{Literature available evidence}

In the preceding section, we show that the eighth-power terms are essential for the adequate description of the dielectric nonlinearity of BaTiO$_3$, especially in the ferroelectric region in the vicinity of the cubic-tetragonal phase transition temperature. Interestingly, this point was not realized for a long time.
The probable reason was that the anharmonic coefficients were in most cases determined by dealing with the properties (e.g., dielectric permittivity,\cite{Drougard1955, Kaczmarek1965} birefringence,\cite{Meyerhofer1958} and phase instability \cite{Merz1953}) of the paraelectric phase using the sixth-power expansion, whereas the dielectric behavior or the polarization of the ferroelectric phase were not considered.
In these cases the contribution of the eighth-power terms is negligible and cannot be detected.
Values of the coefficients reported in these early studies are actually equivalent to our nonlinear coefficients determined from the paraelectric phase only, as explained above.
When the properties of the ferroelectric phase, particularly the very high polarization are involved in determining the anharmonic coefficients of the sixth-power expansion, quite different values for the fourth and sixth order coefficients can be obtained according to our preceding discussion.
Actually this divergency can be found in Gonzalo \textit{et al.}'s work,\cite{Gonzalo1971} where the critical polarization values, at which the paraelectric or the ferroelectric phase loose their stability, have been employed in evaluating the anharmonic coefficients.
The values of their coefficients are nearly the same as our coefficients determined from the ferroelectric phase.
In Table \ref{tab3} we collected values of the coefficients available from the literature, which were determined in the vicinity of $T_C$. The method of determining coefficients is briefly described for each case.
Appreciable difference can be found when comparing Gonzalo \textit{et al.}'s data and those derived from the properties of the paraelectric phase.
Unfortunately this difference of $\sim 100\%$ in magnitude for $\beta$ and $\sim 300\%$ for $\gamma$ were ignored by the authors though it provides a strong signal that the eighth-power terms cannot be neglected.

Additional evidences for the necessity of the eighth-power terms can be found from the linear dielectric responses on crossing the cubic-tetragonal phase transition.
At the first order phase transition temperature $T_C$, under zero bias field, we have
\begin{eqnarray}
\label{Curie conditions}
\Delta G(P_S) & = &\frac{1}{2} \alpha P_S^{2}+\frac{1}{4} \beta P_S^{4}+\frac{1}{6} \gamma P_S^{6}+\frac{1}{8} \delta P_S^{8}=0 \\
E(P_S) & = & \alpha P_S+\beta P_S^{3}+ P_S^{5}+\delta P_S^{7}=0 \\
\epsilon_{c,T}^{-1} & = & \alpha + 3\beta P_S^{2}+ 5 P_S^{4}+ 7 \delta P_S^{6}\\
\epsilon_{c,C}^{-1} & = & \alpha = \alpha_0 (T_C-T_0)
\end{eqnarray}
where $P_S$ is the spontaneous polarization of the tetragonal phase at $T_C$. $\epsilon_{c,C}$ and $\epsilon_{c,T}$ represent the $c$-axis dielectric permittivity of the cubic and the tetragonal phase on crossing the phase transition, respectively. $\alpha_0$=$1/(C \epsilon_0)$ is the temperature derivative of $\alpha$ and $C$, $T_0$ the Curie-Weiss constant and Curie-Weiss temperature, respectively.

Accordingly, two quantities, namely the variation of the dielectric permittivity and the variation of the temperature derivative of the inverse permittivity along $c$-axis on crossing the ferroelectric phase transition, can be analytically derived as
\begin{eqnarray}
\label{Curie variation}
R & = & \frac{\displaystyle \epsilon_{c,C}}{\displaystyle \epsilon_{c,T}} = 4+\frac{\displaystyle \delta P_S^6}{\displaystyle \alpha} \\
R^{\prime} & = & -\frac{\displaystyle \frac{d(1/\epsilon_{c,T})}{dT}}{\displaystyle \frac{d(1/\epsilon_{c,C})}{dT}} = 8+\frac{\displaystyle 18 \delta P_S^2}{\displaystyle 4 \gamma +9 \delta P_S^2}-3 \frac{P_S^2}{\alpha_0} \frac{d \beta}{d T}-5 \frac{P_S^4}{\alpha_0} \frac{d \gamma}{d T}-\cdots
\end{eqnarray}

Using the coefficients values and their temperature derivatives of the existing Landau potentials,\cite{Li2005, Bell1984, Bell2001} we can estimate the numerical values for the two quantities.
Table \ref{tab4} lists the numerical results calculated by the analytical derivations with sixth- and eighth-power expansions.
In case of the sixth-power expansion, namely $\delta=0$, $R=4$ and $R'\sim 8$.
In contrast, the existence of $\delta$ that has a value of the order of that suggested by Li \textit{et al.}, leads to appreciable larger values for both quantities.
Available experimental data are also collected for comparison.
Apparently most of the experimental values are considerable larger than those derived by the sixth-power expansion.
Considering the possible experimental inaccuracy in determining the lattice permittivity in the ferroelectric phase due to the not fully eliminated domain contribution, the real values of $R$ is expected to be even larger.\cite{adiabatic}
The deviations of experimentally determined $R$ and $R^{\prime}$ from those calculated from the sixth-power expansion, affords additional evidence for the essentiality of the eighth-power term.

In summary the necessity of the eighth-power terms in BaTiO$_3$ Landau expansion is demonstrated by our dielectric nonlinearity measurements and literature available data. Keeping in mind that the existence of the eighth-power terms leads to the polarization dependence of the anharmonic coefficients in the sixth-power expansion according to Eq. \eqref{beta'}, the temperature dependence of the quartic coefficients, which was usually determined with the sixth-power expansion,\cite{Drougard1955, Meyerhofer1958} might be logically ascribed to the eighth-power terms as the polarization around $T_C$ is strongly temperature dependent. In next section we will address the temperature dependence of the anharmonic coefficients in the presence of the eighth-power terms and show that the incorporation of the eighth-power terms cannot eliminate the temperature dependence of the fourth order anharmonic coefficients.

\section{TEMPERATURE DEPENDENCE OF THE ANHARMONIC COEFFICIENTS}

\subsection{Temperature dependence of the quartic coefficients}
Though the existence of the non-zero eighth-power terms leads to polarization dependent corrections to the quartic and sextic coefficients, however, these corrections are negligible in the paraelectric phase regime, where the contribution of the anharmonic terms is dominated by the quartic terms.\cite{Wang2006}
Consequently the values determined with the sixth-power expansion of paraelectric properties have a high accuracy. As can be seen from Fig. \ref{fig3}, the quartic coefficient determined by the sixth-power and eighth-power expansion agree well when only paraelectric data are considered.
Therefore we cannot disregard the temperature dependence of $\beta$ suggested by Drougard \cite{Drougard1955} and Meyerhofer \cite{Meyerhofer1958} according to the nonlinear dielectric or birefringence behavior of the paraelectric phase.
A linear temperature dependence was independently revealed for $\beta$ at temperatures above but not far above $T_C$.
Figure \ref{fig4-a} illustrates the quartic coefficient $\beta$ as a function of temperature in the vicinity of $T_C$. An apparent temperature dependence can be found in the paraelectric phase, well agreeing with Drougard and Meyerhofer's results.
Linearly extrapolating to higher temperatures shows that $\beta$ vanishes at around 185$^\circ C$, which, however, is inconsistent with our previous observations of around 165$^\circ C$.\cite{Wang2006}
The difference in the zero cross-over temperature of $\beta$ may be influenced by many factors, e.g. the conductivity, charge accumulating at the oxide-electrode interface, and processing history of the sample, all of which can strongly affect the accuracy of the nonlinear measurements.
It is important, however, that both our previous and present work show qualitatively the same behavior of $\beta$, which is temperature dependent and changes sign at some temperature well above $T_C$.
In the ferroelectric phase, the $\beta$ values determined with the eighth-power expansion and temperature independent $\gamma$ and $\delta$ exhibit a continuous and similar temperature dependence as in paraelectric phase (see Fig. \ref{fig3}).
However this temperature dependence is, by itself, not convincing since we cannot extract sufficient information to verify the temperature dependence of the higher-than-fourth power anharmonic coefficients merely from the dielectric nonlinearity measurements.
The necessity of the temperature dependent anharmonic coefficients will be demonstrated in the next subsection.

The value of another quartic coefficient, $\xi$ [see Eq. \eqref{Gibbs4}], can be evaluated by combining the nonlinear dielectric responses along [001], [011], and [111] orientations of the paraelectric phase.
In this case the anharmonic contributions are dominated by the quartic terms, and the higher-than-fourth-power terms can be omitted.\cite{Wang2006}
The free energy expression can be simplified as
\begin{eqnarray}
\label{Gibbs4}
\triangle G &=& \frac{1}{2} \alpha (P_1^{2}+P_2^{2}+P_3^{2})+\frac{1}{4} \beta (P_1^{4}+P_2^{4}+P_3^{4}) \nonumber \\
& & +\frac{1}{2} \xi (P_1^{2}P_2^{2}+P_2^{2}P_3^{2}+P_3^{2}P_1^{2})-E_1P_1-E_2P_2-E_3P_3
\end{eqnarray}
where $P_1$, $P_2$, $P_3$ and $E_1$, $E_2$, $E_3$ are cartesian components of polarization and electric field, respectively. The nonlinear dielectric response along three orientations can be deduced by successive differentiation:\cite{Belokopytov1995}
\begin{equation}
\label{E4}
E_{hkl}=\alpha P_{hkl}+\beta_{hkl}P_{hkl}^3
\end{equation}
\begin{equation}
\label{lambda4}
\epsilon_{hkl}^{-1}=\alpha+3\beta_{hkl}P_{hkl}^2
\end{equation}
where the effective nonlinear coefficients $\beta_{hkl}$ are linear combinations of $\beta$ and $\xi$:
\begin{eqnarray}
\label{beta_hkl}
\beta_{001}&=&\beta \nonumber \\
\beta_{011}&=&\frac{1}{2}(\beta+\xi) \\
\beta_{111}&=&\frac{1}{3}(\beta+2\xi) \nonumber
\end{eqnarray}
Figure \ref{fig4-b} illustrates the values of $\xi$ as a function of temperature.
The values determined from various combinations agree with each other with an accuracy of 10\% and show consistent temperature dependence over the studied temperature interval.
However, our results determined from dielectric nonlinearity are quite different from those proposed in Ref. \onlinecite{Meyerhofer1958}, where the values of $\xi$ were directly estimated by fitting the electric field dependence of the induced polarization of the paraelectric phase.
Both the value itself and its temperature derivative have opposite signs in comparison with those reported by Meyerhofer\cite{Meyerhofer1958} in the vicinity of $T_C$.
We believe that our positive $\xi$ values are correct for the following reasons. Both [011] and [111] oriented crystals exhibit a decreasing electric field dependence of the dielectric permittivity, suggesting positive values of the effective nonlinear coefficients, $\beta_{011}$ and $\beta_{111}$.
In contrast, a negative $\xi$ definitely results in negative $\beta_{011}$ and $\beta_{111}$, which is opposite to our experimental data.
A negative value of $\xi$ suggested in Ref. \onlinecite{Meyerhofer1958} is actually in contradiction with the birefringence measurements reported in the same paper. \cite{Meyerhofer_contradiction}
Finally, $\xi$ must exhibit a value larger than the absolute value of $\beta$ to fit the positive $\beta_{011}$.
A smaller value of $\xi$, as was implicitly assessed by fitting the tetragonal/orthorhombic phase transition temperature with the sixth-power expansion,\cite{Bell1984, Bell2001, Huibregtse1956} cannot describe the dielectric nonlinearity of [011] crystals in the paraelectric phase.

Therefore, we show that even in the presence of eighth-power terms, which are actually not important for the paraelectric phase, both of the quartic coefficients definitely have to be temperature dependent, at least above $T_C$.
The temperature dependence of the higher-power coefficients, however, cannot be precisely determined merely from dielectric nonlinearity results.
More measurements, e.g. the polarization as a function of electric field, are needed to clarify this point.

\subsection{Temperature dependence of the anharmonic coefficients in tetragonal phase}
In ferroelectric phases below $T_C$, the contributions of the terms higher than quartic are comparable to those of the quartic terms.\cite{Wang2006}
Neither the coefficients themselves nor their temperature dependence can be determined from dielectric nonlinearity in a reasonable accuracy.
Here we propose a simple way to cross-check the necessity of the temperature dependent anharmonic coefficients in tetragonal phase.
Firstly we assume that the Landau coefficients are temperature independent except the quadratic coefficient, $\alpha$, which is linearly temperature dependent following Curie-Weiss law.
The equation of state and $c$-axis permittivity $\epsilon_c$ of the tetragonal phase can be simplified as:
\begin{equation}
\label{E_simple}
E=\alpha_0(T-T_0) P+g(P)
\end{equation}
\begin{equation}
\label{lambda_simple}
\epsilon_c^{-1}= \alpha_0 (T-T_0) + g^{\prime}(P)
\end{equation}
where $g(P)$ summarizes all the contributions from the higher order anharmonic terms.
Under zero electric field, $E=0$, $P=P_S$. Assuming that the only temperature dependence in $g$ comes from $P_S$, one can readily derive
\begin{equation}
\label{dE/dT}
\frac{\mathrm{d}E}{\mathrm{d}T}=0=\alpha_0P_S+[\alpha_0 (T-T_0)+g^{\prime}(P_S)]\frac{\mathrm{d}P_S}{\mathrm{d}T}
\end{equation}
Thus we have
\begin{equation}
\label{g'}
g^{\prime}(P_S)=-\frac{\alpha_0P_S}{\mathrm{d}P_S/\mathrm{d}T}-\alpha_0(T-T_0)
\end{equation}
Combining Eq. \eqref{lambda_simple} and \eqref{g'}, we get
\begin{equation}
\label{epsilon_c}
\epsilon_{r,c}=-\frac{\displaystyle 1}{\displaystyle \alpha_0 \epsilon_0}\frac{\displaystyle d \mathrm{ln} P_S}{\displaystyle dT}=-C\frac{\displaystyle d \mathrm{ln} P_S}{\displaystyle dT}
\end{equation}
where $\epsilon_{r,c}$ represents the relative dielectric permittivity along the direction of the spontaneous polarization $P_S$ in tetragonal phase.
Thus by comparing the quantity $-\epsilon_{r,c}/(d \mathrm{ln} P_S/dT)$ with $C$, we can readily verify the validity of the temperature independent assumption.

In Fig. \ref{fig5} we compare the two quantities as a function of temperature in the tetragonal phase. The experimental data are extracted from Ref. \onlinecite{Merz1953}, where the measurements were performed on identical single domain BaTiO$_3$ crystals.
Appreciable difference ($\sim$ 50\% of $C$) can be found between these two quantities, attesting to the invalidity of the temperature independent assumption.
Namely, in ferroelectric phase, the temperature dependent anharmonic coefficients are indeed necessary for an adequate description of the thermodynamics of BaTiO$_3$.
However, detailed information about the temperature dependence of each coefficient cannot be obtained by this simple method.

\section{IMPROVED THERMODYNAMIC POTENTIAL FOR BaTiO$_3$ SINGLE CRYSTALS}
In the preceding sections we demonstrate that not only the eighth-power terms, but also the temperature dependent anharmonic coefficients (at least temperature dependent quartic coefficients), are essential in BaTiO$_3$ Landau thermodynamic potential.
Hereinafter we propose the following improved potential for this prototype substance by introducing temperature dependent coefficients into the eighth-order expansion.
For convenience we use the same notations as in Ref. \onlinecite{Li2005}:
\begin{eqnarray}
\label{Gibbs_Li}
\triangle G & = & \alpha_1 (P_1^{2}+P_2^{2}+P_3^{2})+ \alpha_{11}(P_1^{4}+P_2^{4}+P_3^{4}) \nonumber \\
& &  +\alpha_{12} (P_1^{2}P_2^{2}+P_2^{2}P_3^{2}+P_3^{2}P_1^{2}) + \alpha_{111} (P_1^{6}+P_2^{6}+P_3^{6}  \nonumber \\
& &  +\alpha_{112} \left[P_1^{2}(P_2^{4}+P_3^{4})+P_2^{2}(P_3^{4}+P_1^{4})+P_3^{2}(P_1^{4}+P_2^{4})\right]  \nonumber \\
& &  +\alpha_{123} P_1^{2}P_2^{2}P_3^{2} +\alpha_{1111} (P_1^{8}+P_2^{8}+P_3^{8})  \nonumber \\
& &  +\alpha_{1112} \left[P_1^{6}(P_2^{2}+P_3^{2})+P_2^{6}(P_3^{2}+P_1^{2})+P_3^{6}(P_1^{2}+P_2^{2})\right]  \nonumber \\
& &  +\alpha_{1122} (P_1^{4}P_2^{4}+P_2^{4}P_3^{4}+P_3^{4}P_1^{4})  \nonumber \\
& &  +\alpha_{1123} (P_1^{4}P_2^{2}P_3^{2}+ P_1^{2}P_2^{4}P_3^{2}+P_1^{2}P_2^{2}P_3^{4}) \nonumber \\
& &  -E_1P_1-E_2P_2-E_3P_3
\end{eqnarray}
where the coefficients $\alpha_1$, $\alpha_{11}$, $\alpha_{12}$, $\alpha_{111}$, and $\alpha_{1111}$ corresponding to the preceding notations $\frac{\displaystyle 1}{\displaystyle 2}\alpha$, $\frac{\displaystyle 1}{\displaystyle 4}\beta$, $\frac{\displaystyle 1}{\displaystyle 2}\xi$, $\frac{\displaystyle 1}{\displaystyle 6}\gamma$, and $\frac{\displaystyle 1}{\displaystyle 8}\delta$, respectively.
The temperature dependent $\alpha_1$, $\alpha_{11}$, and $\alpha_{12}$, are directly determined from our linear and nonlinear dielectric measurements above $T_C$.
The linear dependences are simply extrapolated down to low temperatures.
The temperature dependence of $\alpha_{111}$ is estimated by fitting the spontaneous polarization of the tetragonal phase, with a temperature independent $\alpha_{1111}$ that is evaluated from the dielectric nonlinearity above $T_C$.
A linear dependence is revealed for $\alpha_{111}$ and extrapolated to low temperatures.
The rest of the coefficients are assumed to be temperature independent.
$\alpha_{112}$ and $\alpha_{1112}$ are estimated by fitting the dielectric permittivity perpendicular to the spontaneous polarization, $\epsilon_a$, as a function of temperature in the tetragonal phase.
$\alpha_{1122}$ is evaluated to fulfill the tetragonal/orthorhombic phase transition temperature.
$\alpha_{123}$ and $\alpha_{1123}$ are adjusted to fit the orthorhombic/rhombohedral phase transition temperature and its field dependence.\cite{Fesenko1981}
The numerical values of the nonlinear coefficients described above are listed in Tab. \ref{tab5}. For comparison those of the two existing potentials proposed by Bell \cite{Bell1984} and Li \textit{et al.} \cite{Li2005} are also presented.

In the first glance, our model is just a combination of those proposed by Li \textit{et al.} and Bell.\cite{Bell1984,Li2005}
However it should be kept in mind that
(i) the eighth-power terms used in Li \textit{et al.}'s potential are not just a convenience but an essentiality in eliminating the ``phase sensitivity'' of the anharmonic coefficients of the sixth-power expansion;
(ii) even with the eighth-power expansion, the temperature dependence of the anharmonic coefficients cannot be disregarded, as has been generally demonstrated by the dielectric properties of the tetragonal phase (Fig. \ref{fig5}), and directly revealed by our dielectric measurements for the two quartic coefficients (Fig. \ref{fig4}).
Disregarding these temperature dependence of the quartic coefficients and the eighth order terms leads to an inadequate description of the dielectric behavior, even in a qualitative way.
In contrast, incorporating these changes, a high quality in reproducing the dielectric behavior of BaTiO$_3$ single crystals is achieved, particularly in the vicinity of the cubic-tetragonal phase transition.
Firstly, above $T_C$ our model is capable in quantitatively fitting the dielectric nonlinearity as a function of temperature and crystallographic orientation (see Fig. \ref{fig2} and \ref{fig4}), which in principle cannot be well described by either of the other potentials.
Secondly, in the tetragonal phase the linear dielectric constant and the spontaneous polarization can be precisely reproduced.
Figure \ref{fig6} compares the experimentally measured and the theoretically calculated $1/ \epsilon_{c,r}$ as a function of temperature around the paraelectric/ferroelectric phase transition.
In spite of the great difference existing among the investigated crystals (e.g. the $T_C$ varies from 108 $^\circ C$ for Merz's crystal \cite{Merz1953} to 130 $^\circ C$ of our crystal), the measured dielectric permittivity concide on departing from $T_C$.
The curve calculated with the potential proposed in this work agrees well with the experimental data, whereas those calculated with Li \textit{et al}'s and Bell's coefficients show remarkable divergence with the experimental values in tetragonal phase.
Note that when determining the anharmonic coefficients we have never used the experimental $\epsilon_c$ data measured below $T_C$.

The low temperature ferroelectric properties are also calculated with the three potentials listed in Tab \ref{tab5}.
The three consecutive phase transition temperatures are yielded with the current potential as 404 $K$, 284 $K$, and 189 $K$, respectively.
In Fig. \ref{fig7}, we present the spontaneous polarization along pseudo cubic [001] of BaTiO$_3$ as a function of temperature.
Some of the experimental data available in literature are also presented for comparison.
In the tetragonal phase, the three potentials have similar quality in reproducing the spontaneous polarization, both in its absolute value and in its temperature dependence.
In the orthorhombic phase none of these models can well describe the temperature dependence of the spontaneous polarization. The existing potentials (Li \textit{et al}'s and Bell's) give better estimations of the average value of the spontaneous polarization.
In the rhombohedral phase, however, our potential exhibits a better accuracy than the other two potentials.
Again we emphasize that the experimentally measured polarization values of the rhombohedral phase have not been used in determining the anharmonic coefficients.
As for the dielectric properties of the low temperature ferroelectric phases, experimental data available in literature are rather scattered and sometimes are not self-consistent.\cite{Merz1953, Wemple1968, Huibregtse1956}
Tentatively we measured a poled [111] crystal at cryogenic temperatures.
At 20 $K$ the relative dielectric constant is about 40. Considering the possible incompletely eliminated domain contributions, the real value seems to be even less.
The corresponding calculations give 30, 81, and 68 when using our, Li \textit{et al.}'s and Bell's potentials, respectively.
Apparently the thermodynamic potential proposed in this work exhibits a better accuracy.

\section{CONCLUSIONS}
In conclusion, we demonstrate by analyzing the dielectric nonlinearity that both the eighth-power terms and the temperature dependent anharmonic coefficients are necessary in BaTiO$_3$ Landau phenomenological potential for the adequate description of the thermodynamic behavior of this substance.
Accordingly, an improved Landau potential, characterized by an eighth-order Landau expansion with some temperature dependent anharmonic coefficients, is proposed and compared with the existing potentials.
In general the proposed potential exhibits a higher quality in reproducing the dielectric and ferroelectric properties of this prototype ferroelectric substance.

\clearpage
\newpage
\section*{ACKNOWLEDGEMENTS}
This project was supported in part by the Swiss National Science Foundation. Additional support from
MIND-European net-work on piezoelectricity is gratefully acknowledged. A. I. Sokolov acknowledges the support of
the Russian Foundation for Basic Research under Grants No 07-02-00345, No 04-02-16189 and of Russian Ministry of
Science and Education within Project RNP.2.1.2.7083. The work of I. Lukyanchuk was supported by F6 European
Project "MULTICERAL".

\newpage


\newpage
\begingroup
\begin{table}[h]
\caption{Estimated values of the polarization dependent terms in Eq. \eqref{beta'} and typical values of the corresponding coefficients. The value of $\delta$ is taken from \cite{Li2005}}
\label{tab1}
\begin{tabular}{p{2cm} p{2.5cm} p{2.5cm} p{2cm}}
\hline
\hline
Quantities & Paraelectric & Ferroelectric & Units \\
\hline
$P$ & $\leqslant 0.04$ & $\geqslant 0.16$ & $ (C/m^2) $\\
$\delta P^4 $ & $\leqslant 5 \times 10^4$ & $\geqslant 2 \times 10^8$ & $(Vm^5C^{-3})$\\
$2\delta P^2$ & $\leqslant 2 \times 10^8$ & $\geqslant 1.5 \times 10^{10}$ & $ (Vm^9C^{-5})$\\
$-\beta$ & \multicolumn{2}{c}{ $(6\sim 9) \times 10^8$} & $(Vm^5C^{-3})$ \\
$\gamma$ & \multicolumn{2}{c}{ $(1\sim 2) \times 10^{10}$} & $(Vm^9C^{-5})$\\
\hline
\hline
\end{tabular}
\end{table}
\endgroup

\newpage

\begingroup
\begin{table}
\caption{Anharmonic coefficients determined from the field dependence of dielectric permittivity shown in Fig. \ref{fig2}. The values proposed by Li \textit{et al.} are also listed for comparison.}
\label{tab2}
\begin{tabular}{p{1.5cm} p{4cm} p{3cm} p{3cm} p{3cm}}
\hline
\hline
Order & Phase(s)/Author & $-\beta\mathrm{(Vm^5C^{-3})}$ & $\gamma\mathrm{(Vm^9C^{-5})}$ & $\delta\mathrm{(Vm^{13}C^{-7})}$  \\
\hline
6 & Paraelectric & $8.6\times 10^8$ & $1\times 10^{10}$ \footnotemark & 0 \\
6 & Ferroelectric & $18.7\times 10^8$ & $8.2\times 10^{10}$ & 0 \\
8 & Both & $8.6\times 10^8$ & $0.8\times 10^{10}$ & $3.4\times10^{11}$ \\
8 & Li \textit{et al.} \cite{Li2005} &$8.4\times 10^8$ & $0.77\times 10^{10}$ & $3.0\times10^{11}$ \\
\hline
\hline
\end{tabular}
\footnotetext{The value of $\gamma$ in paraelectric phase cannot be determined only by fitting the electric field dependence of the permittivity. The present value is estimated by combining the dielectric nonlinearity and the position of the critical field $E_C$.}
\end{table}
\endgroup
\clearpage
\newpage

\begingroup
\squeezetable
\begin{table}
\caption{Literature available values of the anharmonic coefficients for BaTiO$_3$ determined with the sixth-power expansion in the vicinity of $T_C$.}
\label{tab3}
\begin{tabular}{p{3cm} p{3cm} p{3cm} p{7cm}}
\hline
\hline
Authors & $-\beta\mathrm{(Vm^5C^{-3})}$ & $\gamma\mathrm{(Vm^9C^{-5})}$ & Principles of the determination  \\
\hline
Merz \cite{Merz1953} & $5.4\times 10^8$ & $1.7\times 10^{10}$ & Instability of the paraelectric phase under bias field \\
Drougard \cite{Drougard1955} & $(4\sim8)\times 10^8$ & NA & Polarization dependence of the dielectric permittivity in paraelectric phase \\
Meyerhofer \cite{Meyerhofer1958} &  $(7\sim10)\times 10^8$ & $ (2\sim2.5) \times 10^{10}$ & Field dependence of the polarization in paraelectric phase\\
Kaczmarek \cite{Kaczmarek1965} & $6\times 10^8$ & NA &  Field dependence of the dielectric permittivity in paraelectric phase\\
Gonzalo \cite{Gonzalo1971} & $(15\sim20)\times 10^8$ & $(5\sim10)\times 10^{10}$ & Instability of the paraelectric and ferroelectric phase above $T_C$\footnotemark \\
\hline
\hline
\end{tabular}
\footnotetext{In this work, the critical polarization (or the polarization at the destabilizing electric field) of either phases, were determined from double hysteresis loops. These values in practice strongly depend on the measuring and sample conditions, e.g. the switching frequency and electrode/crystal interface natures. Even though, the differences are appreciable when comparing with those determined from paraelectric properties.}
\end{table}
\endgroup
\clearpage
\newpage
\begingroup
\begin{table}
\caption{Analytical calculated and experimentally determined values of $R$ and $R^{\prime}$ for BaTiO$_3$ single crystals. The anharmonic coefficients and their temperature derivatives are taken from literature \cite{Li2005, Bell1984, Bell2001}. The spontaneous polarization are taken from literature \cite{Merz1953}.}
\label{tab4}
\begin{tabular}{p{2cm} p{2cm} p{2cm} l}
\hline
\hline
Quantities & 6th-power expansion & 8th-power expansion & Experimental data  \\
\hline
$R$ & 4 & 5.3 & 6.2\footnotemark[2], 7.3\footnotemark[3], 3.3\footnotemark[4], 6.1\footnotemark[5], 5.2\footnotemark[6] \\
$R^{\prime}$ & 7.9\footnotemark[1] & 9.3 & 12\footnotemark[7], 9.4\footnotemark[6] \\
\hline \hline
\footnotetext[1]{As can be seen that the contributions by the temperature dependence of the anharmonic coefficients to $R^{\prime}$ are quite weak and in an opposite direction in comparison to that by the eighth-power term.}
\footnotetext[2]{Ref. \cite{Merz1953}}
\footnotetext[3]{Ref. \cite{Drougard1954}}
\footnotetext[4]{Ref. \cite{Meyerhofer1958}}
\footnotetext[5]{Ref. \cite{Johnson1965}}
\footnotetext[6]{This work}
\footnotetext[7]{Ref. \cite{Draegert1971}}
\end{tabular}
\end{table}
\endgroup
\clearpage
\newpage
\begingroup
\squeezetable
\begin{table}
\caption{Coefficients of BaTiO$_3$ Landau thermodynamic potential in Eq. \eqref{Gibbs_Li}, where $T$ is temperature in K.}
\label{tab5}
\begin{tabular}{p{2cm} p{4cm} p{3.5cm} p{4cm} l}
\hline
\hline
Coefficients & This work & Li \textit{et al.} \footnotemark[1] & Bell and Cross \footnotemark[2] & Units \\
\hline
$\alpha_1$ & $3.61 \times 10^5(T-391)$ & $4.124 \times 10^5(T-388)$ & $3.34 \times 10^5(T-381)$ & $VmC^{-1}$ \\
$\alpha_{11}$ & $-1.83 \times 10^9+4 \times 10^6T$ & $-2.097 \times 10^8$ & $-2.045 \times 10^9+4.69 \times 10^6T$ & $Vm^5C^{-3}$ \\
$\alpha_{12}$ & $-2.24 \times 10^9+6.7 \times 10^6T$ & $7.974 \times 10^8$ & $3.23 \times 10^8$ & $Vm^5C^{-3}$ \\
$\alpha_{111}$ & $1.39 \times 10^{10}-3.2 \times 10^7T$ & $1.294 \times 10^9$ & $2.445 \times 10^{10}-5.52 \times 10^7T$ & $Vm^9C^{-5}$ \\
$\alpha_{112}$ & $-2.2 \times 10^9$ & $-1.950 \times 10^9$ & $4.47 \times 10^9$ & $Vm^9C^{-5}$ \\
$\alpha_{123}$ & $5.51 \times 10^{10}$ & $-2.500 \times 10^9$ & $4.91 \times 10^9$ & $Vm^9C^{-5}$ \\
$\alpha_{1111}$ & $4.84 \times 10^{10}$ & $3.863 \times 10^{10}$ & $0$ & $Vm^{13}C^{-7}$ \\
$\alpha_{1112}$ & $2.53 \times 10^{11}$ & $2.529 \times 10^{10}$ & $0$ & $Vm^{13}C^{-7}$ \\
$\alpha_{1122}$ & $2.80 \times 10^{11}$ & $1.637 \times 10^{10}$ & $0$ & $Vm^{13}C^{-7}$ \\
$\alpha_{1123}$ & $9.35 \times 10^{10}$ & $1.367 \times 10^{10}$ & $0$ & $Vm^{13}C^{-7}$ \\
\hline
\hline
\footnotetext[1]{Ref. \cite{Li2005}}
\footnotetext[2]{Ref. \cite{Bell1984, Bell2001}}
\end{tabular}
\end{table}
\endgroup
\clearpage
\newpage

\clearpage
\newpage
\begin{figure}[h]
\begin{center}
\caption{Schematic of the phase transition induced by an electrical field applied along [001] axis above \textit{T$_C$} for BaTiO$_3$ single crystal. In paraelectric phase the polarization varies with field below $P_C$, while in ferroelectric phase the polarization sits in the interval above $P_0$.}
\label{double hysteresis loop}
\end{center}
\end{figure}

\begin{figure}[h]
\begin{center}
\caption{(Color online) Field dependence of the dielectric permittivity for BaTiO$_3$ single crystal at 135 $^\circ C$. The open squares represent the measured values, and the dotted line (blue) is to guide the eyes. Remarkably different $\beta$ are obtained for either phase when fitting with Eq. \eqref{E6} and \eqref{lambda6}, as listed in Table \ref{tab1}. The dashed (red) and solid (green) lines are calculated with the coefficients derived from the paraelectric and ferroelectric data, respectively. The ferroelectric coefficients have a serious problem in describing the dielectric properties of the paraelectric phase, and vice versa. The inset shows the much better fit quality (black dashed-dotted line) of an eighth-order expansion for both phases.}
\label{fig2}
\end{center}
\end{figure}

\begin{figure}[h]
\begin{center}
\caption{(Color online) The quartic coefficient determined with sixth and eighth-power expansions in the vicinity of $T_C$. Phase dependent $\beta^{\prime}$ values are derived when using sixth-power expansion, whereas phase independent $\beta$ can be obtained with the eighth-power expansion. The $\beta$ values below $T_C$ are estimated by fitting the dielectric nonlinearity with constant $\gamma$ and $\delta$. However this is less convincing since the temperature dependence of $\gamma$ and $\delta$ have been arbitrarily ommitted.}
 \label{fig3}
\end{center}
\end{figure}

\begin{figure}[h]
\begin{center}

\caption{(Color online) Temperature dependence of the quartic coefficients $\beta$ (a) and $\xi$ (b). The values of $\xi$ are determined by combining the dielectric nonlinearity measurements along [001], [011] and [111] orientations, and the values and temperature derivatives calculated from three combinations agree well with each other.}
\label{fig4}
\end{center}
\end{figure}

\begin{figure}[h]
\begin{center}
\caption{(Color online) Divergency between the quantity $-\frac{\displaystyle \epsilon_{r,c}}{\displaystyle d \mathrm{ln} P_S/ dT}$ and $C$ in the ferroelectric tetragonal phase. The values of dielectric permittivity and spontaneous polarization are extracted from literature \cite{Merz1953}.}
\label{fig5}
\end{center}
\end{figure}

\begin{figure}[h]
\begin{center}
\caption{(Color online) Inverse dielectric permittivity along the spontaneous polarization in the vicinity of $T_C$, a comparison between the calculations and the experimental data. For convenience the phase transition temperatures are normalized to 120$^\circ C$. The scattered dark marks are experimentally measured by different researchers on different crystals:  ``+'' by Drougard \cite{Drougard1954}, ``$\bigcirc$'' by Merz \cite{Merz1953}, and ``$\Box$'' in this work. The dashed (blue), dash-dotted (green), and solid (red) lines are calculated with Li \textit{et al.}'s, Bell's and the newly-proposed potential in this work, respectively. Note that despite the great difference in the source sample, the permittivity exhibits tremendous consistence on departing from $T_C$. The new potential has a much better quality in fitting the experimental data than the two existing potentials do.}
\label{fig6}
\end{center}
\end{figure}

\begin{figure}[h]
\begin{center}
\caption{(Color online) Spontaneous polarization along pseudo cubic [001] of BaTiO$_3$, a comparison between the calculations and the experimental data. The scattered marks represent the experimental data:  ``$\Box$'', ``$\blacklozenge$'', and ``$\bigtriangleup$'' are extracted from literature \cite{Merz1953, Wemple1968, Wieder1955}, respectively. The dashed (dark), dash-dotted (red) and solid (blue) lines are calculated with  Li \textit{et al.}'s, Bell's and the newly-proposed potential in this work, respectively. The polarization values of the rhombohedral phase are exactly reproduced by the improved thermodynamic potential.}
\label{fig7}
\end{center}
\end{figure}


\begin{thebibliography}{999}

\bibitem{Vanderbilt2001} D. Vanderbilt, and M. H. Cohen, Phys. Rev. B \textbf{63}, 94108, (2001).
\bibitem{Gufan2002} I. A. Sergienko, Yu. M. Gufan, and S. Urazhdin, Phys. Rev. B \textbf{65}, 144104, (2002).
\bibitem{Noheda1999} B. Noheda, D. E. Cox, G. Shirane, J. A. Ganzalo, L. E. Cross and S-E. Park, Appl. Phys. Lett. \textbf(74), 2059 (1999).
\bibitem{Li2005} Y. L. Li, L. E. Cross, and L. Q. Chen, J. Appl. Phys. \textbf{98}, 64101, (2005).
\bibitem{Bell1984} A. J. Bell and L. E. Cross, Ferroelectric. \textbf{59}, 197, (1984).
\bibitem{Bell2001} A. J. Bell, J. Appl. Phys. \textbf{89}, 3907, (2001).
\bibitem{Vaks1970} V. G. Vaks, Sov. Phys. -JETP \textbf{31}, 161, (1970).
\bibitem{Sokolov2002} A. I. Sokolov, and A. K. Tagantsev, Pis'ma v ZhETF, \textbf{75}, 483, (2002).
\bibitem{Iniguez2001} J. Iniguez, S. Ivantchev, J. M. Perez-Mato, and A. Garcia, Phys. Rev. B \textbf{63}, 144103, (2001).
\bibitem{Wang2006} Y. L. Wang, A. K. Tagantsev, D. Damjanovic, N. Setter, V. K. Yarmarkin, and A. I. Sokolov, Phys. Rev. B \textbf{73}, 132103 (2006).
\bibitem{Triebwasser1960} S. Triebwasser, Phys. Rev. \textbf{118}, 100, (1960).
\bibitem{Drougard1955} M. E. Drougard, R. Landauer, and D. R. Young, Phys. Rev. \textbf{98}, 1010, (1955).
\bibitem{Kaczmarek1965} F. Kaczmarek and J. Pietrzak, Acta Physica Polonica \textbf{XXVII}, 335, (1964).
\bibitem{Meyerhofer1958} D. Meyerhofer, Phys. Rev. \textbf{112}, 413, (1958).
\bibitem{Merz1953} W. J. Merz, Phys. Rev. \textbf{91}, 513, (1953).
\bibitem{Gonzalo1971} J. A. Gonzalo and J. M. Rivera, Ferroelectric. \textbf{2}, 31, (1971).
\bibitem{Draegert1971} D. A. Draegert and S. Singh, Solid State Commun. \textbf{9}, 595, (1971).
\bibitem{adiabatic} The adiabatic correction to the dielectric permittivity measured by ac method, which might affect the value of $R$, is negligible for BaTiO$_3$, according to Ref. \cite{Drougard1955}.
\bibitem{Drougard1954} M. E. Drougard and D. R. Young, Phys. Rev. \textbf{95}, 1152, (1954).
\bibitem{Johnson1965} C. J. Johnson, Appl. Phys. Lett. \textbf{7}, 221, (1965).
\bibitem{Belokopytov1995} G. V. Belokopytov, Ferroelectric. \textbf{168}, 69, (1995).
\bibitem{Meyerhofer_contradiction} In literature \cite{Meyerhofer1958}, the second quartic coefficient $\xi$, was determined by fitting the P-E curves along [001] and [011] directions in the paraelectric phase. Neglecting the high-power terms, $P_{hkl}=\kappa_{hkl} E_{hkl}+\lambda_{hkl} E_{hkl}^3$, and -$\beta_{hkl}=\lambda_{hkl}/\epsilon^4$. As determined $\lambda_{001}$ and $\lambda_{011}$ both have positive values above $T_C$. Keeping in mind that the birefringence are proportional to the square of the polarization, $\Delta n \propto P^2 = \kappa^2 E^2+2 \kappa \lambda E^4+\cdots$, one would expect that the nonlinear coefficient $d_2$ in $\Delta n=d_1E^2+d_2E^4+\cdots$ have the same sign as $\lambda$ for both directions as $\kappa_{001}=\kappa_{011}$. However the birefringence measurements evidenced positive $d_{2[001]}$ whereas negative $d_{2[011]}$ over the same temperature interval, contradicting to that predicted by $\lambda$ values.
\bibitem{Huibregtse1956} E. J. Huibregtse and D. R. Young, Phys. Rev. \textbf{103}, 1705 (1956).
\bibitem{Fesenko1981} O. E. Fesenko and V. S. Popov, Ferroelectric. \textbf{37}, 729 (1981).
\bibitem{Wemple1968} S. H. Wemple, M. Didomenico, Jr. and I. Camlibel, J. Phys. Chem. Solids \textbf{29}, 1797, (1968).
\bibitem{Wieder1955} H. H. Wieder, Phys. Rev. \textbf{99}, 1161, (1955).
\end{thebibliography}
 \end{document}